\documentclass[twocolumn,superscriptaddress,prl]{revtex4}

\usepackage[latin9]{inputenc}
\usepackage{color}
\usepackage{verbatim}
\usepackage{amsmath}
\usepackage{amssymb}
\usepackage{graphicx}
\usepackage{esint}
\usepackage{hyperref}

\makeatletter
\@ifundefined{textcolor}{}
{%
 \definecolor{BLACK}{gray}{0}
 \definecolor{WHITE}{gray}{1}
 \definecolor{RED}{rgb}{1,0,0}
 \definecolor{GREEN}{rgb}{0,1,0}
 \definecolor{BLUE}{rgb}{0,0,1}
 \definecolor{CYAN}{cmyk}{1,0,0,0}
 \definecolor{MAGENTA}{cmyk}{0,1,0,0}
 \definecolor{YELLOW}{cmyk}{0,0,1,0}
}

\usepackage{mathrsfs}

\draft
\begin{document}

\title{Chiral Topological Orders in an Optical Raman Lattice}

\author{Xiong-Jun Liu}
\affiliation{Department of Physics, Hong Kong University of Science and Technology, Clear Water Bay, Hong Kong, China}
\affiliation{International Center for Quantum Materials and School of Physics, Peking University, Beijing 100871, China}
\affiliation{Department of Physics, Massachusetts Institute of Technology, Cambridge, Massachusetts 02139, USA}
\author{Zheng-Xin Liu}
\affiliation{Institute for Advanced Study, Tsinghua University, Beijing 100084, P. R. China}
\author{K. T. Law}
\affiliation{Department of Physics, Hong Kong University of Science and Technology, Clear Water Bay, Hong Kong, China}
\author{W. Vincent Liu}
\affiliation{Department of Physics and Astronomy, University of Pittsburgh, Pittsburgh, Pennsylvania 15260, USA}
\author{T. K. Ng}
\affiliation{Department of Physics, Hong Kong University of Science and Technology, Clear Water Bay, Hong Kong, China}

\begin{abstract}
We find an optical Raman lattice without spin-orbit coupling showing chiral topological orders for cold atoms. Two incident plane-wave lasers are applied to generate simultaneously a double-well square lattice and periodic Raman couplings, the latter of which drive the nearest-neighbor hopping and create a staggered flux pattern across the lattice. Such a minimal setup is can yield the quantum anomalous Hall effect in the single particle regime, while in the interacting regime it achieves the $J_1$-$J_2$-$K$ model with all parameters controllable, which supports a chiral spin liquid phase. We further show that heating in the present optical Raman lattice is reduced by more than one order of magnitude compared with the conventional laser-assisted tunneling schemes. This suggests that the predicted topological states be well reachable with the current experimental capability.
\end{abstract}
\date{\today}
\maketitle

Generation of synthetic gauge fields for cold atoms opens a new direction in the study of exotic topological states beyond natural conditions. Two different scientific paths have been followed in the experiment to create synthetic gauge fields via optical means. One is to adopt Raman couplings between different internal hyperfine levels (atomic spins)~\cite{Ruseckas2005,Osterloh2005,Liu1,Zhu1,Stanescu2007,Liu2009}, which has been recently used in experiments to generate synthetic spin-orbit (SO) coupling for cold atoms~\cite{Lin,MIT,Wang}. Another is to adopt laser-assisted hopping between neighboring lattice sites without spin flip, which can generate U(1) fluxes by imprinting the phases of Raman lasers into the hopping matrix elements~\cite{Bloch2011,Bloch2013,Ketterle2013a,Ketterle2013b,Struck2013}. Compared with the technique using spin-flip Raman couplings, the latter strategy can be achieved with far-detuned lasers, and therefore can avoid the spontaneous decay of excited states.

Realization of a gapped (insulating) topological state typically necessitates an optical lattice and synthetic gauge fields which satisfy proper conditions~\cite{Bloch2013,Ketterle2013a,Ketterle2013b,Struck2013,Congjun2008,Sato2009,Liu2010,Goldman2010,Goldman2012,Hauke,Liu2013a,Vincent2013,Wong2013, Chuanwei2013,Wei2013,Luming,Cooper}. In the conventional techniques, the optical lattice and gauge fields are generated through different atom-laser couplings. In such cases the topological regimes are achieved with careful manipulations of parameters, which might be challenging for the experimental observation. Recently, it was proposed that creations of the optical lattice and SO couplings can be integrated through the same standing-wave lasers, and this new technique can have explicit advantages in realizing topological phases with minimal setups and without complicated manipulations~\cite{Liu2013a,Liu2014}. Nevertheless, generating SO couplings requires near-resonant light which heats up the system by spontaneous emission~\cite{Lin,MIT,Wang}. A possible resolution of this difficulty is to consider lanthanide atoms which can have less heating due to large fine structure splitting and narrow natural linewidth in the excited levels~\cite{Zhai2013}.

In this letter, we introduce the model of optical Raman lattice without SO coupling to observe chiral topological phases for cold atoms. The setup includes a double-well square lattice and periodic Raman couplings generated simultaneously through two incident plane-wave beams. We show that this scheme can naturally realize chiral topological phases without fine tunings, and has essential advantages in the experimental observation including the minimized heating and full controllability in parameters.

\begin{figure}[ht]
\includegraphics[width=0.9\columnwidth]{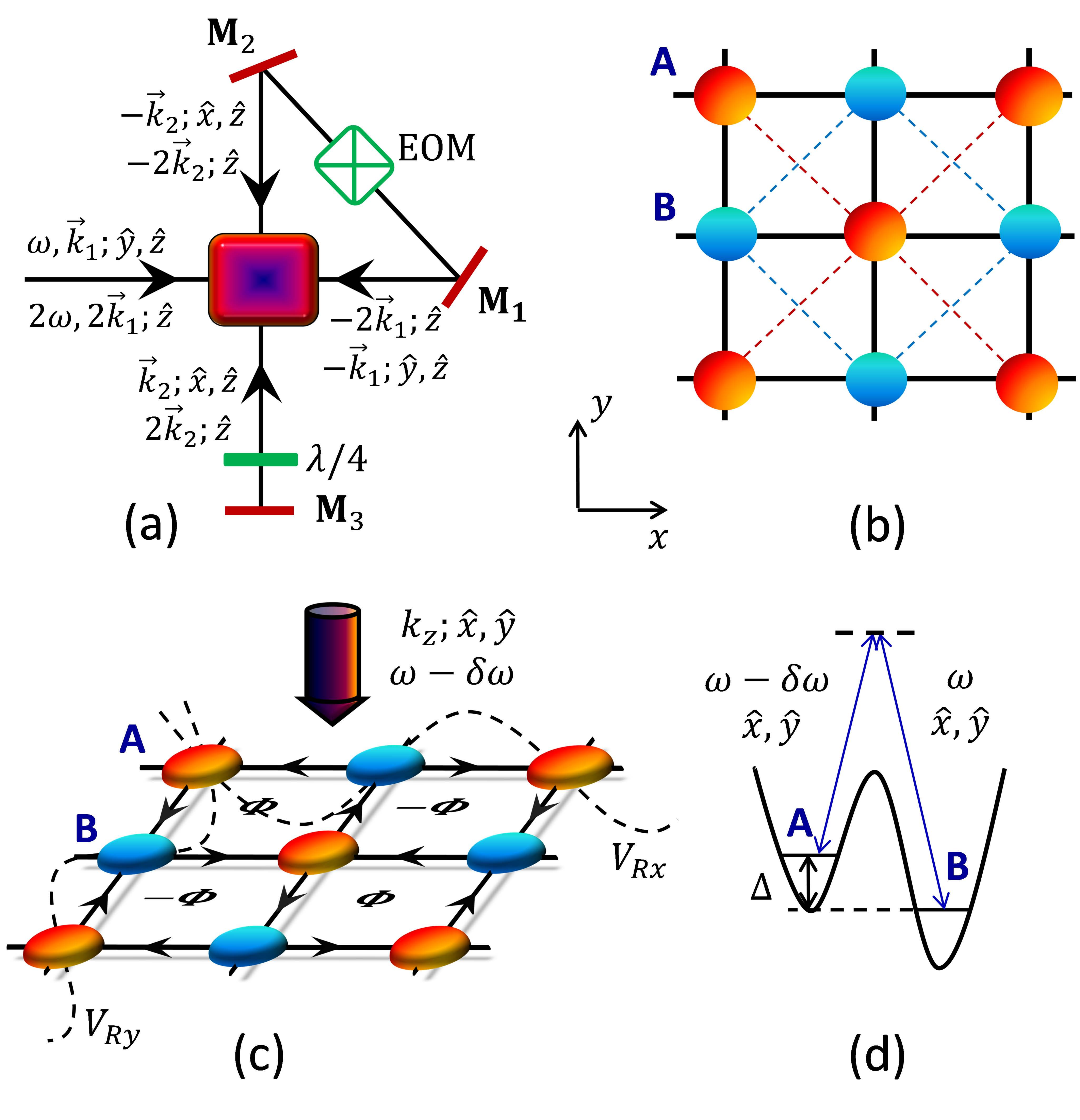}
\caption{(Color online) (a) The laser beam incident from $x$ direction has nonzero polarization components along $y$ and $z$ axes, with the $\hat z$ polarization field having two components of wave-vectors $k_1$ and $2k_1$. With the reflections of the three mirrors $M_j$ ($j=1,2,3$), a double-well lattice is generated and can be controlled by the $1/4$-wave plate and electro-optic modulator (EOM)~\cite{SI}; (b) The created double-well lattice has an energy off-set between $A$ and $B$ sublattices; (c) An additional linearly-polarized running laser beam, with polarization components in $x$ and $y$ directions, is applied along $z$ direction. This beam, together with the laser components used to create square lattice, can induce spatially periodic Raman couplings ($V_{Rx}, V_{Ry}$) between $A$ and $B$ sites, as illustrated in (d). The Raman transitions also generate a staggered flux pattern simultaneously for the nearest-neighbor hopping.}\label{lattice}
\end{figure}
We first introduce the generation of a 2D double well square lattice depicted in
Fig. \ref{lattice}(a-b), with an onsite energy difference $\Delta$ between $A$ and $B$ sites. Based on the experiments by NIST group~\cite{Porto}, this lattice configuration can be realized via an incident plane-wave laser beam which has both nonzero in-plane and out-of-plane linearly polarized components, and with the assistance of three mirrors [see Fig.~\ref{lattice}~(a)]. The total electric field of the incident laser beam can be described as $\bold E(x)=E_0(\cos\alpha\hat y+\sin\alpha\hat z)e^{i(k_1x-\omega t)}+\tilde E_0\hat z e^{i2(k_1x-\omega t)}$, where the polarization angle $\alpha$ determines the magnitudes of in-plane and out-of-plane polarization components with wave-vector $k_1$, and the our-of-plane polarized field has another component with its wave vector ($2k_1$) being twice of that of the former ones. Note that all the components of the incident beam can be generated from a single laser source through an optical frequency doubler, with which one can control the ratio of the field strengths $|E_0/\tilde E_0|$ in the experiment~\cite{EOM}. With the reflection by mirrors the $\hat y$-polarization component of the incident laser beam changes to be $\hat x$-polarization component when the laser beam propagates along $\pm y$ direction, while the $\hat z$-polarization components do not change for the entire optical paths. The in-plane polarized components ($\hat x$ and $\hat y$ polarization components) generate the standing wave as $\bold E_{xy}=2E_0\cos\alpha\bigr[\cos(k_1x)\hat y+\cos(k_1y)\hat x\bigr]e^{-i\omega t}$. Here we have neglected all irrelevant constant phase factors which have no effect on our results~\cite{SI}.
On the other hand, the out-of-plane light components can interfere and generate the standing wave as $E_{z}\hat z=4\tilde E_0\sin[k_1(x+y)]\sin[k_1(x-y)]e^{-i2\omega t}\hat z+4E_0\sin\alpha \sin\bigr[\frac{k_1}{2}(x+y)\bigr]\cos\bigr[\frac{k_1}{2}(x-y)\bigr]e^{-i\omega t}\hat z$. Furthermore, we consider here the {\it blue-detuned} optical dipole transitions. The total lattice potential reads
\begin{eqnarray}\label{eqn:lattice1}
V_{\rm sq}(x,y)&=&V_0\bigr[\cos^2(k_1x)+\cos^2(k_1y)\bigr]\nonumber\\
&&+\tilde V_0\sin^2[k_1(x+y)]\sin^2[k_1(x-y)]\nonumber\\
&&+\Delta\sin2\bigr[\frac{k_1}{2}(x+y)\bigr]\cos^2\bigr[\frac{k_1}{2}(x-y)\bigr].
\end{eqnarray}
The amplitudes are taken that $\Delta<V_0$, and the above potential describes a double-well square lattice illustrated in Fig.~\ref{lattice} (b), with the staggered onsite energy offset $\Delta$ between $A$ and $B$ sites well controlled by the polarization angle $\alpha$. When $\Delta$ is large compared with the bare hopping couplings between neighboring $s$-orbitals of the $A$ and $B$ sites, the effective tunneling between them is suppressed, while the diagonal $AA/BB$ hoppings (denoted by $t'_{A/B}$) are allowed along dashed lines in Fig.~\ref{lattice} (b). The second term in Eq.~\eqref{eqn:lattice1} reduces the difference in height of the barriers along the $AB$-bond and the diagonal ($AA/BB$) directions. Thus it can enhance  $t'_{A/B}$ relative to the hopping coupling between $A$ and $B$ sites, providing vast tunability in parameters.

The tunneling between neighboring $A$ and $B$ sites (denoted by $t_{\vec i\vec j}$) can be restored by two-photon Raman couplings. A key ingredient of the present scheme is that the in-plane blue-detuned laser beam which generates the square lattice also takes part in the generation of Raman couplings. For this we apply an additional plane-wave laser beam with frequency $\omega-\delta\omega$ ($\delta\omega\approx\Delta$), propagating along the perpendicular $z$ direction and having linear polarization components along $x$ and $y$ axes [Fig.~\ref{lattice} (c)]. This beam is described by $\tilde{\bold E}_{xy}(z)=E_1(e^{i\phi_x}\hat x+e^{i\phi_y}\hat y)e^{ik_zz}$, where $\phi_{x/y}$ is the initial phase of the $x/y$-axis polarization component. With the assistance of both $\bold E_{xy}$ and $\tilde{\bold E}_{xy}$, two independent Raman couplings are induced by the $\hat x$- and $\hat y$-polarization components, respectively [Fig.~\ref{lattice} (d)]. In particular, the $\hat x$ $(\hat y)$-components of the lights $\bold E_{xy}$ and $\tilde{\bold E}_{xy}$ generate the Raman potential $V_{Rx}$ ($V_{Ry}$) which takes the form
\begin{eqnarray}\label{eqn:Raman1}
V_{Rx(Ry)}=V_R\cos[k_1x(y)]e^{i\delta\omega t+i\phi_{y(x)}}+c.c.,
\end{eqnarray}
where the amplitude $V_R\propto E_0E_1\cos\alpha$. We shall see below that a finite magnitude of $\phi_x-\phi_y$, controllable in experiment, gives rise to a nonzero staggered flux pattern for the square lattice, as illustrated in Fig.~\ref{lattice} (c).

From Eqs.~\eqref{eqn:lattice1}-\eqref{eqn:Raman1} we can see that the zeros of $V_{Rx, Ry}$ are located at the lattice-site centers, which implies that the Raman potentials are parity odd relative to each lattice-site center [Fig.~\ref{lattice} (c)]. With this key property the present blue-detuned optical Raman lattice can naturally realize topological states and exhibit essential advantages in minimize heating effects for experimental studies. The symmetry properties of $s$-orbitals and $V_{Rx, Ry}$ lead to two important consequences. First, the Raman potential $V_{Rx}$ ($V_{Ry}$) only induces the nearest-neighbor hopping along $x$ ($y$) direction. The hopping along $x/y$ axis is associated with a phase $\phi_{y/x}$ ($-\phi_{y/x}$) if the hopping is toward (away from) $B$ sites. In experiment, one can set that $\phi_x-\phi_y=2\phi_0$, which is equivalent to put that $\phi_x=-\phi_y=\phi_0$. Then the hopping along the directions depicted by arrows in Fig.~\ref{lattice} (c) acquires a phase $\phi_0$, resulting in a staggered flux pattern with the flux $\Phi=4\phi_0$. Secondly, the hopping from one site to its leftward (upward) neighboring site has an additional minus sign relative to the hopping to its rightward (downward) neighboring site. It is important that all these interesting properties are obtained automatically by using the two incident beams without complicated fine tunings.

\begin{figure}[ht]
\includegraphics[width=1.0\columnwidth]{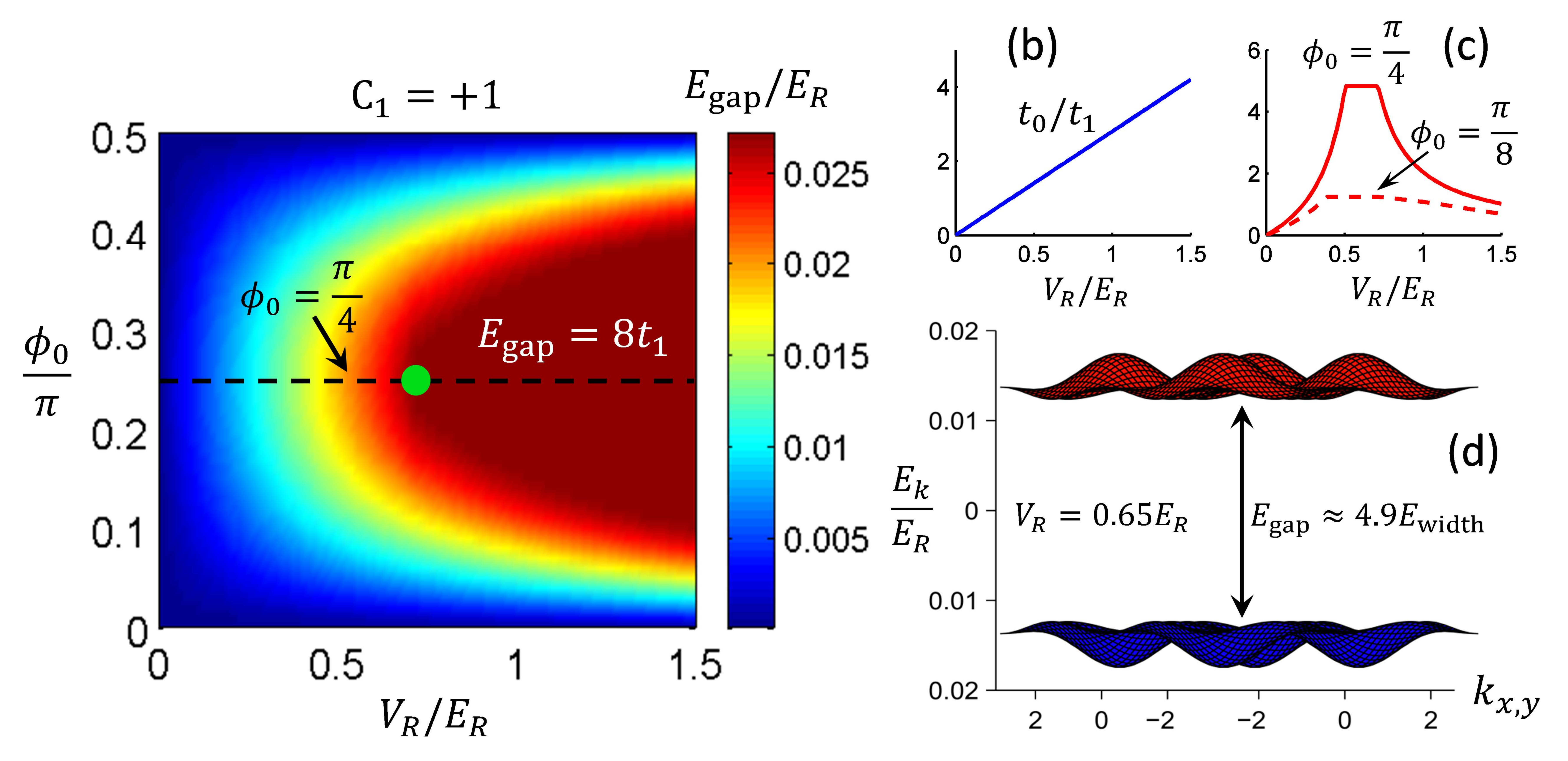}
\caption{(Color online) (a) The band gap $E_{\rm gap}$ versus Raman potential amplitude $V_R$ (in units of the recoil energy $E_R$) and the hopping phase $\phi_0$; (b) The ratio $t_0/t_1$ with $t_1=0.5(t'_A+t'_B)$ and (c) the ratio $E_{\rm gap}/E_{\rm width}$ (blue curve) as functions of $V_R$. The phase $\phi_0=\pi/4$ ($\pi/8$) for solid (dashed) curves; (d) The bulk spectrum with $V_R=0.65E_R$ and $\phi_0=\pi/4$. Other parameters are $V_0/E_R=4, \tilde V_0/E_R=1$, and $\delta\omega=\Delta=0.6E_R$. For $^{87}$Rb atoms using $\lambda=532$nm yields $E_R\approx2\pi\times8$kHz, and $t_1\simeq2\pi\times27$Hz. The Chern number $C_1=+1$.}\label{fig2}
\end{figure}

Now we give the tight-binding Hamiltonian (details are shown in Supplementary Material~\cite{SI}). Based on the previous analysis we have that the nearest-neighbor hopping coefficients (excluding the hopping phases) satisfy
$t_{\vec i,\vec i\pm1_x}=\pm(-1)^{i_x}t_0,
t_{\vec i,\vec i\pm1_y}=\mp(-1)^{i_x}t_0$,
with $t_0=V_{R}\int d^2\bold r\psi_{B,s}^{(0,0)}(\bold r)\sin(k_1x)\psi_{A,s}^{(1,0)}(\bold r)$ and $\psi_{\mu,s}^{(\vec j)}(\bold r)$ ($\mu=A,B$) the $s$-orbital wave function at the $\vec j$-th site. The staggered sign factor $(-1)^{i_x}$ is due to the staggered position distribution of $A$ and $B$ sites, and can be absorbed by redefining the annihilation operator of $B$ sites to be $c_{B,\vec j}=e^{i\pi x_j/a}c_{B,\vec j}$, with $a$ the lattice constant. In terms of the new basis, the diagonal hopping coefficient for the $B$ sites reverses sign $t'_B\rightarrow-t'_B$~\cite{SI}.
The tight-binding model can now be obtained directly and in $\bold k$ space the Bloch Hamiltonian reads
${\cal H}(\bold k)=-2t_0\cos\phi_0(\sin k_xa+\sin k_{y}a)\sigma_x-2t_0\sin\phi_0(\sin k_xa-\sin k_{y}a)\sigma_y+\bigr[m_z-2(t'_A+t'_B)\cos k_{x}a\cos k_{y}a\bigr]\sigma_z$, with the Zeeman term $m_z=(\Delta-\delta\omega)/2$ and $\sigma_{x,y,z}$ the Pauli matrices.
It is interesting that without Zeeman and diagonal hopping terms, i.e. if $m_z=t'_{A,B}=0$, the above Hamiltonian would describe massless Dirac fermions with two independent Dirac points at $\bold\Lambda_1=(0,0)$ and $\bold\Lambda_2=(0,\pi)$. The bulk is gapped when $m_z\neq2(t'_A+t'_B)$ and $\phi_0\neq n\pi/2$ with $n$ being integer. The quantum anomalous Hall (QAH) phases are obtained for $|m_z|<2(t'_A+t'_B)$, with the first Chern number $C_1=\mbox{sgn}(\phi_0)$ ($0<|\phi_0|<\pi/2$), and trivial regime results for $|m_z|>2(t'_A+t'_B)$.
Fig.~\ref{fig2} provides the numerical estimate with $V_0/E_R=4, \tilde V_0/E_R=1$, $\Delta=0.6E_R$, and the recoil energy $E_R\approx2\pi\times8$kHz using $\lambda=2\pi/k_1=532$nm for $^{87}$Rb atoms, which gives that $t'_{A,B}\simeq2\pi\times27$Hz. By setting $\phi_0=\pi/4$ and $m_z=0$, the bulk gap $E_{\rm gap}=4(t'_A+t'_B)\approx2\pi\times0.21$kHz when $t_0>t'_A+t'_B$ for $V_R>0.71E_R$ [Fig.~\ref{fig2}(a)]. Fig.~\ref{fig2} (b-d) show a large ratio ($\sim4.9$) between the band gap and bandwidth $E_{\rm width}$ in the range from $t_0=0.7(t'_A+t'_B)$ at $V_R\simeq0.51E_R$ to $t_0= t'_A+t'_B$ at $V_R\simeq0.7E_R$. It is noteworthy that a large gap-bandwidth ratio can enable the study of correlated topological states like the fractional QAH effect~\cite{FQAHE} in the interacting regime.

Detection of the QAH insulating phase can be carried out with several different measurement strategies in the edge~\cite{Liu2010,Goldman2012} and bulk~\cite{Alba,Demler,WangLei,Goldman2013,Dongling}. In particular, it was proposed recently that the topology of a QAH insulator can be determined by measuring Bloch eigenstates at only two or four highly symmetric points of the first Brillouin zone~\cite{Liu2013b}. In Supplementary Material we show that this minimal-measurement approach can be applied to the present system with a nontrivial generalization~\cite{SI}.
It can be seen that ${\cal H}(\bold k)$ is formally invariant under the 2D transformation $P=\sigma_z\otimes R_{\rm 2D}$, where $\sigma_z$ is the ``parity operator" acting on the pseudospin space, and $ R_{\rm 2D}$ sends $(k_{x},k_{y})\rightarrow(-k_{x},-k_{y})$. At the highly symmetric points $\{\bold\Lambda_i\}=\{(0,0),(0,\pi),(\pi,0),(\pi,\pi)\}$, with only $\Lambda_{1,2}$ being independent for the physical Hamiltonian, the Bloch states are also parity eigenstates with $\sigma_z|u_{\pm}(\bold \Lambda_i)\rangle=\xi^{(\pm)}_i|u_{\pm}(\bold \Lambda_i)\rangle$, with $\xi^{(\pm)}_i=+1$ or $-1$. To measure the parity eigenvalues $\xi^{(\pm)}_i$ one can measure the signs of pseudospin polarization, i.e. the population difference of atoms between $A$ and $B$ sublattice sites, which can be measured with {\it in situ} imaging. For a cloud of bosons condensed at some momentum $\bold k$, the pseudospin polarization is defined by $p_s=(N_A-N_B)/(N_A+N_B)$, where $N_{A/B}$ represents the number of atoms in the $A/B$ sublattice. The topology of the QAH insulator is determined by the following invariant
\begin{eqnarray}\label{eqn:detection}
(-1)^{\nu}&=&\mbox{sgn}\bigr[p_s(\bold\Lambda_1)\bigr]\mbox{sgn}\bigr[p_s(\bold\Lambda_2)\bigr].
\end{eqnarray}
The topological phase corresponds to $\nu=-1$ for $|m_z|<2(t'_A+t'_B)$, and the trivial phase corresponds to $\nu=0$ for $|m_z|>2(t'_A+t'_B)$. The detection can be carried out with a pseudospin-resolved Bloch oscillation~\cite{Liu2013b}. With an external force applied along the $x$ direction, the momentum of the condensate evolves along the direction from $\bold\Lambda_1=(0,0)$ to $\bold\Lambda_2=(0,\pi)$. For the topological phase, the pseudospin polarization $p_s(\tau)$ reverses sign from $\mbox{sgn}[p_s(0)]=-1$ to $\mbox{sgn}[p_s(\tau)]=+1$ at half Bloch time $\tau=T_B/2$ and returns to $\mbox{sgn}[p_s(\tau)]=-1$ at $\tau=T_B$. On the other hand, in the trivial regime, the sign of $p_s(\tau)$ keeps unchanged during the Bloch oscillation. Note that only qualitative measurements at the two symmetric momenta are needed for the experimental detection.

The experimental feasibility of observing topological states, especially the correlated topological states, crucially depends on heating effects in the realization. While the laser-assisted tunneling scheme without spin flip does not suffer from large spontaneous decay from excited states, the heating can be induced by onsite two-photon Raman transitions which do not drive neighboring-site hopping but convert the energy difference between two Raman photons to mechanical energy of the lattice system~\cite{Ketterle2013a,Ketterle2013b}. Note that in the present optical Raman lattice, due to the antisymmetry of the Raman potentials the intraband scattering ($s\leftrightarrow s$ bands) is forbidden, and only the interband scattering ($s\leftrightarrow p$ bands) can heat the system. This distinguishes essentially from the conventional schemes which apply plane-wave and red-detuned Raman beams and have both inter- and intraband onsite transitions~\cite{Bloch2011,Bloch2013,Ketterle2013a,Ketterle2013b,Struck2013}. Denote by $\Gamma_{\rm OR}$ the heating rate of the optical Raman lattice system, and $\Gamma^{\rm min}_{\rm CO}$ the minimum heating rate in the conventional schemes. We can show the following relation (Supplementary Material)
\begin{eqnarray}\label{eqn:lifetime}
\Gamma_{\rm OR}\simeq\frac{1}{16}\frac{\Delta}{E_{sp}-\Delta}\Gamma^{\rm min}_{\rm CO},
\end{eqnarray}
where $E_{sp}=2(V_0E_R)^{1/2}$ is the $s$-$p$ band gap. In the typical regime with $\Delta\ll E_{sp}$ we have $\Gamma_{\rm OR}\ll\Gamma^{\rm min}_{\rm CO}$, which shows that the present optical Raman lattice has much less heating than that in the conventional schemes. In particular, with the parameter regime used in Fig.~\ref{fig2}, one finds that $\Gamma_{\rm OR}\simeq0.03\Gamma^{\rm min}_{\rm CO}$ and for $V_R=0.5E_R$ the life time of the optical Raman lattice $\tau\simeq1.67$s~\cite{SI}, which is extremely long enough for realistic experiments.

\begin{figure}[ht]
\includegraphics[width=1.0\columnwidth]{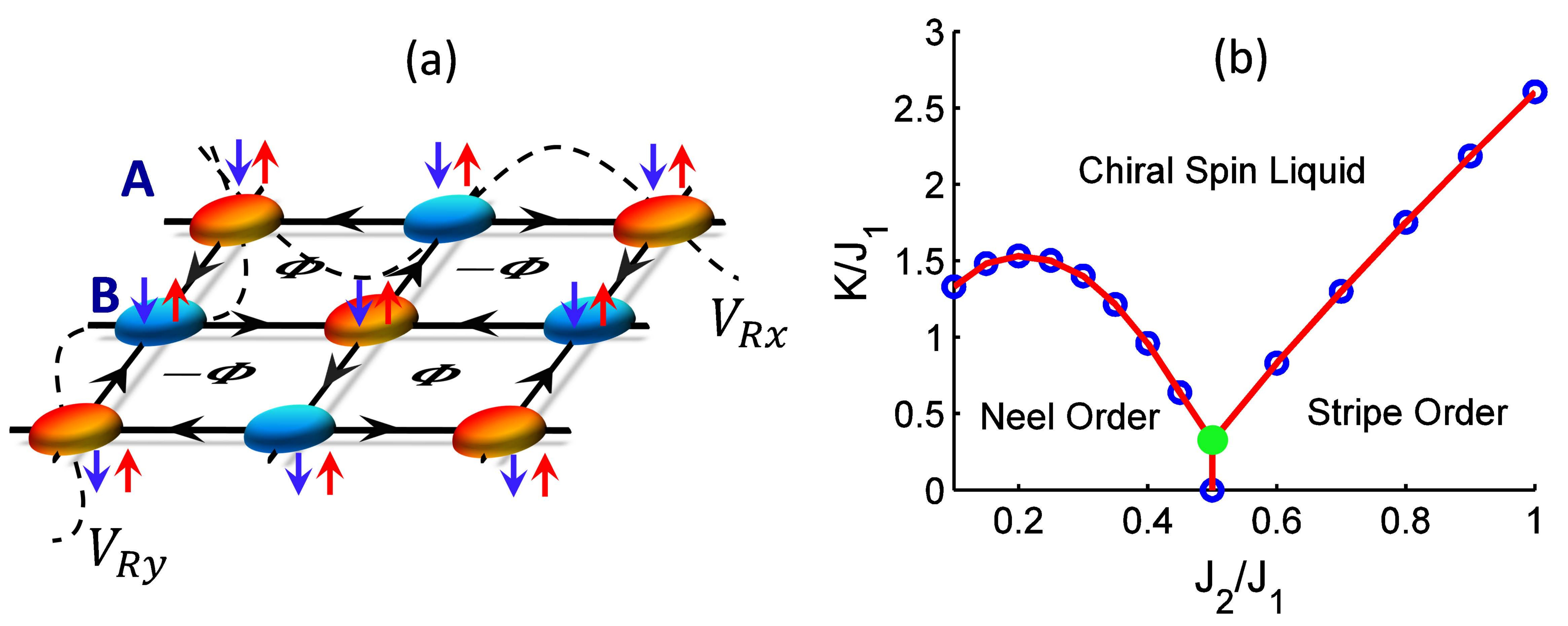}
\caption{(Color online) (a) Sketch of the spin-$1/2$ two copy version of the QAH system. The system is driven into the $J_1$-$J_2$-$K$ spin model with large repulsive Hubbard-$U$. (b) The phase diagram versus $J_2/J_1$ and $K/J_1$. A large region of chiral spin liquid phase is obtained due to the three-spin interactions. The green point corresponds to $J_1=2J_2\approx3.2K$.}\label{phasediagram}
\end{figure}

Finally we turn to the realization of the $J_1$-$J_2$-$K$ model and show it has a large parameter region to support the highly-sought-after chiral spin liquid (CSL) phase~\cite{Kalmeyer,Wen1989,Patrick}. For this we consider a spin-$1/2$ two copy version of the QAH model together with repulsive Fermi Hubbard interaction $H_{int}=\sum_iUn_{i\uparrow}n_{i\downarrow}$ [Fig.~\ref{phasediagram} (a)]. In the single-particle regime each spin species forms a QAH insulator with the same Chern number, while in the large-$U$ regime, the double occupancy of each site will be fully suppressed, and the system becomes a Mott insulator. We can then derive an effective spin-model by considering the perturbation expansion with respect to $t_0/U, t'_\mu/U$, with $t_0,t'_\mu$ being small compared with $U$~\cite{MacDonald1988,Sen}. To reflect the broken time-reversal symmetry, we should at least keep the terms up to third order of $t_0/U,t'_\mu/U$, and then reach the following effective Hamiltonian for the spin degree of freedom~\cite{SI}
\begin{eqnarray}\label{eqn:spinmodel1}
H_{\rm eff}&=&\sum_{\langle i,j\rangle}J_1\bold S_i\cdot\bold S_j+\sum_{\langle\langle i,j\rangle\rangle}J_2\bold S_i\cdot\bold S_j\nonumber\\
&&+\sum_{i,j,k\in\bigtriangleup}K\sin(\phi_{ijk})\bold S_i\cdot(\bold S_j\times\bold S_k),
\end{eqnarray}
where $J_1=4t_0^2/U, J_2=4t_1^2/U$ with $t_1=t'_A\approx t'_B$, $K=24t_0^2t'_1/U^2$, and $\phi_{ijk}$ is the Aharonov-Bohm phase acquired by hopping through a closed triangular loop. It is clear that the third order $K$-term emerges due to the time-reversal-symmetry breaking. The summation in the third term means that each set of $(i,j,k)$ consists of a minimum triangular. It can be verified that $\phi_{ijk}=\pi/2$ when $\phi_0=\pi/4$. In this case all spins experience a uniform magnetic field and the spin system respects the emergent translational symmetry which, however, is not respected by the original free fermion system. The magnitudes of $J_{1,2}$ are fully controllable by tuning $t_{1,2}$ through $\tilde V_0$ and Raman potentials.

We solve the spinon mean-field phase diagram, as shown in Fig.~\ref{phasediagram} (b). It can be read that three different phases are clearly dominated by different interacting terms in $H_{\rm eff}$. The antiferromagnetic (Neel) or stripe order is obtained when the $J_1$- or $J_2$-term dominates. In the stripe phase the staggered spin order exists only in the $x$ or $y$ direction. On the other hand, when the three-spin interactions ($K$-terms) dominate, the CSL phase results~\cite{Motrunich,Katsura2010,Cirac2013,Sheng13}. In this regime, no symmetry-breaking order exists and the spin degree of freedom is captured by the bosonic $\nu=1/2$ Laughlin
state which has bulk semion excitations and chiral gapless spinons in the edge~\cite{Laughlin}. When $K=0$, the transition between Neel and stripe orders occurs at $J_2=J_1/2$, where the system becomes most frustrated. Then increasing $K$ above the point $J_1=2J_2\approx3.2K$ can soon drive and stabilize the CSL phase~\cite{SI}. Taking $V_0=4\tilde V_0=4E_R,V_R=0.52E_R$ and $U=7.6E_{\rm width}$, we find $J_1\approx2J_2\approx2\pi\times19$Hz and $K\approx0.53J_1$, which is in the CSL phase region. Note that while the spinon mean-field calculation only shows qualitative results of the predicted phases, recent numerical simulation using density matrix renormalization group method also confirms the CSL phase in a similar spin model~\cite{Ludwig2014}. More advanced investigations of the current $J_1$-$J_2$-$K$ model will be presented in the next publication.

In conclusion, we have introduced the model of optical Raman lattice without SO coupling to observe chiral topological phases for cold atoms. We predict the QAH effect with a large gap-bandwidth ratio in the single-particle regime, and in the interacting regime we realize the $J_1$-$J_2$-$K$ model which supports the chiral spin liquid phase. The minimized heating in our scheme and vast tunability in parameters imply high feasibility for the observation of both the single-particle and strongly correlated topological states. Generalization of the present optical Raman lattice scheme to other situations, e.g. the high-orbital bands, 3D systems, and more exotic lattice configurations, shall realize different classes of topological states which might even have no prior analogue in solids. Especially, the correlation effects on such topological phases should be particularly interesting. This work opens a broad avenue in both theory and experiment for the studies of exotic topological states with cold atoms.

We appreciate the discussions with Patrick A. Lee, Randy Hulet, Andreas Hemmerich, Chong Wang, and Hong-Hao Tu. We particularly thank Wujie Huang and Colin Kennedy for helpful comments and critical reading of the manuscript. We thank HKRGC for support through DAG12SC01, Grants No. 602813, No. 605512, and No. HKUST3/CRF/13G. ZXL is supported by NSFC 11204149 and Tsinghua University Initiative Scientific Research Program. WVL is supported by AFOSR (FA9550-12-1-0079), ARO (W911NF-11-1-0230), DARPA OLE Program through ARO and the Charles E. Kaufman Foundation of the Pittsburgh Foundation.


\noindent

\onecolumngrid

\renewcommand{\thesection}{S-\arabic{section}}
\renewcommand{\theequation}{S\arabic{equation}}
\setcounter{equation}{0}  
\renewcommand{\thefigure}{S\arabic{figure}}
\setcounter{figure}{0}  

\section*{\Large\bf Supplementary Information}
In this Supplementary Material we provides the details of deriving the tight-binding Hamiltonian, detection of quantum anomalous Hall states, heating effects, $J_1$-$J_2$-$K$ model, and chiral spin liquid phase.

\section{Tight-Binding Model}

\subsection{Optical Raman lattice}

As described in Fig.~1 of the main text, the electric field of the in-plane incident laser beam is $\bold E(x)=E_0(\cos\alpha\hat y+\sin\alpha\hat z)e^{i(k_1x-\omega t)}+\tilde E_0\hat z e^{i2(k_1x-\omega t)}$. The initial relative phase between the light components of frequencies $\omega$ and $2\omega$ is irrelevant for the present study and is neglected. The in-plane polarized components ($\hat x$ and $\hat y$ polarization components) generate the standing waves $\bold E_{xy}=2E_0\cos\alpha\bigr[\cos(k_1x-\theta-\phi+\delta\phi)\hat y+\cos(k_1y+\theta)\hat x\bigr]e^{i(-\omega t+\theta+\phi-\delta\phi)}$.
Here $\theta$ is the phase acquired through the path from mirror $M_3$ to lattice center, $\phi$ represents the phase acquired by the laser beam propagating along the path from lattice center to the mirror $M_1$, then to $M_2$, and finally to the lattice center again (refer to Fig.~1 of the main text), and $\delta\phi$ is the phase tuned through the electric-optic modulator. For our purpose we set that $\delta\phi=\pi$.
On the other hand, the out-of-plane light components generates the standing wave as $E_{z}\hat z=4\tilde E_0\cos[k_1(x+y)-\phi+\frac{\delta\phi}{2}]\cos[k_1(x-y)-\phi-2\theta+\frac{\delta\phi}{2}]e^{i2(-\omega t+\theta+\phi-\frac{\delta\phi}{2})}\hat z+4E_0\sin\alpha\cos\bigr[\frac{k_1}{2}(x+y)-\frac{\phi-\delta\phi}{2}\bigr]\cos\bigr[\frac{k_1}{2}(x-y)-\frac{\phi-\delta\phi}{2}-\theta-\frac{\pi}{2}\bigr]e^{i(-\omega t+\theta+\phi-\delta\phi+\frac{\pi}{2})}\hat z$, where the additional $\pi/2$-phase shift in the later term is due to the $1/4$-wave plate for the $\hat z$-component light with frequency $\omega$ placed in the path from mirror $M_3$ to lattice center [Fig.~1 (a) of the main text]. Note that there is no interference between the two components with frequencies $\omega$ and $2\omega$.
It can be verified that the magnitudes of the phases ($\theta,\phi-\delta\phi$) only lead to global shift of the lattice, and therefore are irrelevant to our present study (they also do not affect the relative phase $\phi_x-\phi_y$ relating to the Raman potentials $V_{Rx}$ and $V_{Ry}$). We then set these phase factors as zero to facilitate the description.

Together with the periodic Raman potentials induced through both the in-plane laser and the one propagating in $z$ direction and having frequency $\omega-\delta\omega$, the total effective Hamiltonian for the optical Raman lattice is given by
\begin{eqnarray}\label{eqn:SIHamiltonian1}
H=\frac{p_x^2+p_y^2}{2m}+V_{\rm sq}(x,y)+\tilde V_{Rx}(x,t)+\tilde V_{Ry}(y,t),
\end{eqnarray}
where $m$ is atom mass, the double-well square lattice potential $V_{\rm sq}(x,y)$, the time-dependent Raman potentials $V_{Rx}(x,t)$ and $V_{Ry}(y,t)$ take the forms
\begin{eqnarray}\label{eqn:SIlattice1}
V_{\rm sq}(x,y)&=&V_0\bigr[\cos^2(k_1x)+\cos^2(k_1y)\bigr]+\tilde V_0\sin^2[k_1(x+y)]\sin^2[k_1(x-y)]+V_1\sin^2\bigr[\frac{k_1}{2}(x+y)\bigr]\cos^2\bigr[\frac{k_1}{2}(x-y)\bigr],\\
V_{Rx}(x,t)&=&2V_R\cos(\delta\omega t-\phi_y)\cos(k_1x), \ V_{Ry}(y,t)=2V_R\cos(\delta\omega t-\phi_x)\cos(k_1y).
\end{eqnarray}
The third term in $V_{\rm sq}(x,y)$ leads to an onsite energy offset $\Delta=V_1$ between $A$ and $B$ sites, with $V_1$ being small compared with $V_0$. The second term with amplitude $\tilde V_0$ reduces the difference in height of the barriers along the $AB$-bond and the diagonal ($AA/BB$) directions. Thus it can enhance the diagonal tunneling relative to the hopping coupling between $A$ and $B$ sites, providing vast tunability in parameters. The neighboring hoppings between $A$ and $B$ sites are restored when $\delta\omega\approx\Delta$. Here we consider only the $s$-orbitals $\psi^{(\vec j)}_{\mu,s}(\bold r)$ ($\mu=A,B$), which are of even parity.

\subsection{Laser-assisted hopping and tight-binding Hamiltonian}

To study the laser-assisted hopping, we examine the symmetry properties of the square lattice potential and Raman potentials $V_{Rx, Ry}$. It can be seen that the zeros of $V_{Rx, Ry}$ are located at the lattice-site centers, implying that the Raman potentials are parity odd relative to each lattice-site center. This configuration is stable against phase fluctuations of the applied lasers, since the periodic properties of the Raman potentials and the square lattice are determined by the same standing-wave electric field $\bold E_{xy}$ generated by the in-plane incident laser beam. The phase fluctuations only leads to global shifts of the lattice and the Raman potentials simultaneously, without changing their relative spatial configuration. For the $s$-orbital bands, we have the following properties.

(i) The Raman potentials cannot induce on-site couplings for the $s$-orbitals, but can induce hopping transitions between neighboring $A$ and $B$ sites.

(ii) The Raman potential $V_{Rx}$ ($V_{Ry}$) induces the hopping along $x$ ($y$) direction, but cannot induce the hopping along $y$ ($x$) direction.

(iii) The hopping along $x/y$ axis is associated with a phase $\phi_{y/x}$ ($-\phi_{y/x}$), when the hopping is toward (away from) $B$ sites. In experiment, one can readily set that $\phi_x-\phi_y=2\phi_0\neq0$, which is equivalent to put that $\phi_x=-\phi_y=\phi_0$, and then the hopping along the directions depicted by arrows in Fig.~1 (c) of the main text acquires a phase $\phi_0$. This leads to a staggered flux pattern with the flux $|\Phi|=4\phi_0$ in each square plaquette.

(iv) Due to the odd parity of Raman potentials, the hopping from one site to its leftward (upward) neighboring site has an additional minus sign relative to the hopping to its rightward (downward) neighboring site.

With these results in mind, we can obtain the $s$-band tight-binding Hamiltonian in the following form
\begin{eqnarray}
H=-\sum_{\langle\vec i,\vec j\rangle}t_{\vec i\vec j}\bigr(\cos\phi_0+i\nu_{\vec i\vec j}\sin\phi_0\bigr) c_{B,\vec i}^\dag c_{A,\vec j}-\sum_{\langle\langle\vec i,\vec j\rangle\rangle}\sum_{\mu=A,B}t'_{\mu,\vec i\vec j} c_{\mu,\vec i}^\dag c_{\mu,\vec j}+m_z\sum_{\vec i}(n_{\vec i,A}-n_{\vec i,B}),
\end{eqnarray}
where $n_{\vec i,\mu}= c^\dag_{\mu,\vec j} c_{\mu,\vec j}$, with $c^\dag_{\mu,\vec j}$ and $ c_{\mu,\vec j}$ the creation and annihilation operator, the factor $\nu_{\vec i\vec j}=1$ ($-1$) for hopping along (opposite to) the marked direction in Fig.~1 (c) of the main text, and the Zeeman term $m_z=(\Delta-\delta\omega)/2$. The nearest-neighbor and diagonal hopping coefficients (excluding the hopping phases), $t_{\vec i\vec j}$ and $t'_{\mu,\vec i\vec j}$, are given by
\begin{eqnarray}
t_{\vec i,\vec i\pm1_x}&=&2V_{R}\frac{1}{T}\int_0^T dt\cos(\delta\omega t)e^{-i\delta\omega t}\int d^2\bold r\psi_{B,s}^{(i_x,i_y)}(\bold r)\cos(k_1x)\psi_{A,s}^{(i_x\pm1,i_y)}(\bold r),\\
t_{\vec i,\vec i\pm1_y}&=&2V_{R}\frac{1}{T}\int_0^T dt\cos(\delta\omega t)e^{-i\delta\omega t}\int d^2\bold r\psi_{B,s}^{(i_x,i_y)}(\bold r)\cos(k_1y)\psi_{A,s}^{(i_x,i_y\pm1)}(\bold r),\\
t'_{\mu,\vec i,\vec j}&=&\int d^2\bold r\psi_{\mu,s}^{(i_x,i_y)}(\bold r)\bigr[\frac{p_x^2+p_y^2}{2m}+V(\bold r)\bigr]\psi_{\mu,s}^{(i_x\pm1,i_y\pm1)}(\bold r),
\end{eqnarray}
where $T=2\pi/\delta\omega$.
It is easy to verify that $t_{\vec i,\vec i\pm1_x}=\pm(-1)^{i_x}t_0,
t_{\vec i,\vec i\pm1_y}=\mp(-1)^{i_x}t_0,
t'_{\mu,\vec i,\vec j}=t'_\mu$,
with $t_0=V_{R}\int d^2\bold r\psi_{B,s}^{(0,0)}(\bold r)\cos(k_1x)\psi_{A,s}^{(1,0)}(\bold r)$ and $t'_\mu=\int d^2\bold r\psi_{\mu,s}^{(0,0)}(\bold r)\bigr[\frac{p_x^2+p_y^2}{2m}+V_{\rm sq}(\bold r)\bigr]\psi_{\mu,s}^{(1,1)}(\bold r)$ ($\mu=A,B$). The staggered sign $(-1)^{i_x}$ can be absorbed by redefining the operator of $B$ sites to be $c_{B,\vec j}=e^{i\pi x_j/a}c_{B,\vec j}$, where $a$ is the lattice constant. Then the tight-binding Hamiltonian can be rewritten as
\begin{eqnarray}\label{eqn:SItightbinding2}
H&=&-\sum_{\vec j}t_0\bigr[(\cos\phi_0-i\sin\phi_0)(\hat c_{B,\vec j+1_x}^\dag\hat c_{A,\vec j}-\hat c_{B,\vec j-1_x}^\dag\hat c_{A,\vec j})+{\rm H.c.}\bigr]-\nonumber\\
&&-\sum_{\vec j}t_0\bigr[(\cos\phi_0+i\sin\phi_0)(\hat c_{B,\vec j+1_y}^\dag\hat c_{A,\vec j}-\hat c_{B,\vec j-1_y}^\dag\hat c_{A,\vec j})+{\rm H.c.}\bigr]-\nonumber\\
&&-\sum_{\langle\langle \vec i,\vec j\rangle\rangle}(t'_A\hat c_{A,\vec i}^\dag\hat c_{A,\vec j}-t'_B\hat c_{B,\vec i}^\dag\hat c_{B,\vec j})+m_z\sum_{\vec i}(n_{\vec i,a}-n_{\vec i,b}).
\end{eqnarray}
Thus in terms of the new basis, the diagonal hopping coefficient for the $B$ sites reverse sign $t'_B\rightarrow-t'_B$. This property will result in quantum anomalous Hall (QAH) effect. Note that the next-order hopping couplings ($AA/BB$ hopping along $x$ and $y$ directions) contribute to a small kinetic term to the Hamiltonian, in the form of $\epsilon(\bold k)\hat I$, with $\hat I$ the unit matrix. This term does not affect the topological phase and thus is neglected.

\section{Quantum anomalous Hall states}

\subsection{Quantum anomalous Hall insulator with large bulk gap}

We transfer the tight-binding Hamiltonian~\eqref{eqn:SItightbinding2} into $\bold k$ space $H=\sum_{\bold k}\hat {\mathcal C}^{\dag}(\bold
k)\mathcal{H}(\bold k)\hat {\mathcal C}(\bold k)$ with $\hat
{\mathcal C}(\bold k)=(\hat c_{A}(\bold k), \hat c_{B}(\bold k))^T$ and obtain
\begin{eqnarray}\label{eqn:SIBlochHamiltonian1}
{\cal H}(\bold k)=d_x(\bold k)\sigma_x+d_y(\bold k)\sigma_y+d_z(\bold k)\sigma_z,
\end{eqnarray}
where the coefficients are
\begin{eqnarray}
d_x=-2t_0\cos\phi_0(\sin k_xa+\sin k_ya), \
d_y=-2t_0\sin\phi_0(\sin k_xa-\sin k_ya),\
d_z=m_z-2(t'_A+t'_B)\cos k_xa\cos k_ya.\nonumber
\end{eqnarray}
When $m_z\neq2(t'_A+t'_B)$ and $\phi_0\neq n\pi/2$ with $n\in\mathbb{Z}$, the bulk is gapped. Note that for $m_z\gg2(t'_A+t'_B)$, the system is trivial insulator. Reducing $m_z$ to be $m_z=2(t'_A+t'_B)$ closes the bulk gap at $\bold k=(0,0)$, and further reducing it reopens the gap. The low-energy physics around $\bold k=(0,0)$ is captured by a massive Dirac Hamiltonian with the mass changing sign when $m_z$ varies from $m_z\gtrsim2(t'_A+t'_B)$ to $m_z\lesssim2(t'_A+t'_B)$. Then the Chern number changes by $1$. Moreover, if tuning $m_z$ down to $-2(t'_A+t'_B)$, another gap-closing occurs at $\bold k=(0,\pi)$, implying that topological phase is obtained with $|m_z|<2(t'_A+t'_B)$, otherwise the phase is trivial. The quantitative calculation of Chern number is $C_1=\frac{1}{4\pi}\int dk_xdk_y\bold n\cdot(\partial_{k_x}\bold n\times\partial_{k_y}\bold n)$, where $\bold n=(d_x,d_y,d_z)/|\vec d(\bold k)|$ with $|\vec d(\bold k)|=(d_x^2+d_y^2+d_z^2)^{1/2}$ represents the unit vector field in the spherical surface. It is easy to show that $C_1$ changes sign when $\phi_0$ varies from $0<\phi_0<\pi/2$ to $\pi/2<\phi<\pi$. We then conclude that $C_1=\mbox{sgn}(\phi_0)$ ($0<|\phi_0|<\pi/2$) for the topological regime.

\begin{figure}[ht]
\includegraphics[width=0.6\columnwidth]{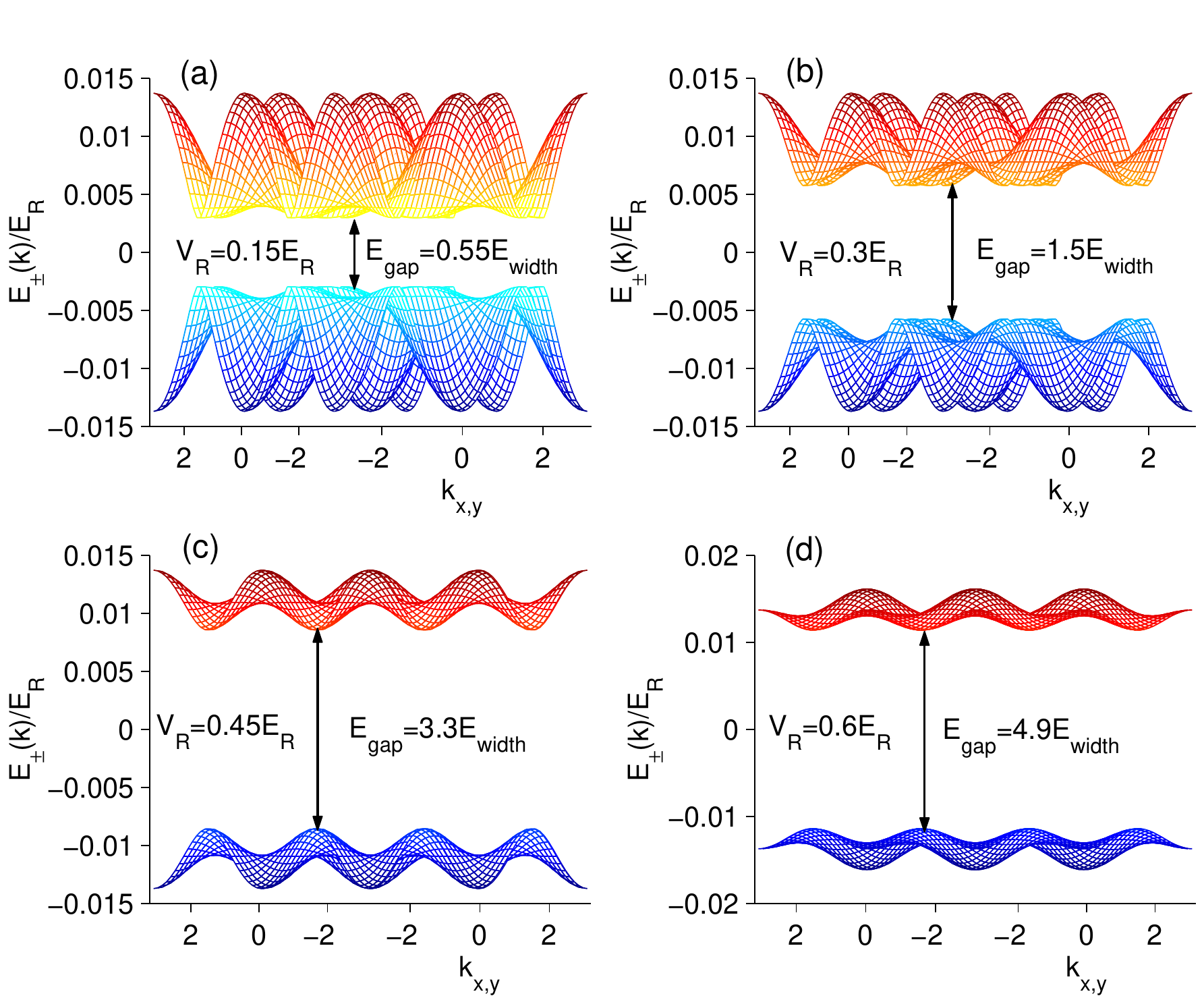}
\caption{(Color online)  The bulk band spectra $E_\pm(\bold k)$ with different Raman potential amplitudes. (a) $V_R=0.15E_R$, and in this case one has $t_0/t_1\approx0.42$, with $t_1=(t'_A+t'_B)/2$; (b) $V_R=0.3E_R$ and $t_0/t_1\approx0.83$; (c) $V_R=0.45E_R$ and $t_0/t_1\approx1.25$; (d) $V_R=0.6E_R$ and $t_0/t_1\approx1.7$.}\label{spectrum}
\end{figure}

We provide the numerical estimate with $V_0/E_R=4, \tilde V_0/E_R=1$, $\Delta=0.6E_R$, and the recoil energy $E_R\approx2\pi\times8$kHz using $\lambda=2\pi/k_1=532$nm for $^{87}$Rb atoms, which gives that $t'_{A,B}\simeq2\pi\times27$Hz. By setting $\phi_0=\pi/4$ and considering the resonant Raman process with $m_z=0$, the bulk gap $E_{\rm gap}=4(t'_A+t'_B)\approx2\pi\times0.21$kHz when $t_0>t'_A+t'_B$ for $V_R>0.71E_R$. A large ratio ($\sim4.9$) between the band gap and bandwidth $E_{\rm width}$ is obtained in the range from $t_0=0.7(t'_A+t'_B)$ at $V_R\simeq0.51E_R$ to $t_0= t'_A+t'_B$ at $V_R\simeq0.7E_R$. In Fig.~\ref{spectrum} the bulk spectra are plotted with different Raman potential amplitudes. It is noteworthy that a large gap-bandwidth ratio can enable the study of correlated topological states including the fractional QAH effect in the strongly interacting regime.

\subsection{Detection}

The simplest approach to detect the topology of a QAH insulator is to measure the Bloch eigenstates at only two or four highly symmetric points in the first Brillouin zone, as proposed in Ref.~\cite{SILiu2013}. In the previous work it was shown that this approach is valid for QAH insulators which satisfy the inversion symmetry defined by $P=\hat P\otimes\hat R_{\rm 2D}$, where $\hat R_{\rm 2D}$ is a 2D spatial inversion operator transforming the Bravais lattice vector $\bold R\rightarrow-\bold R$ and $\hat P$ is a parity operator acting on the (pseudo)spin space. In the present lattice system the unit cell is doubled relative to the original square lattice, and from the Hamiltonian~\eqref{eqn:SIBlochHamiltonian1} one can check that no parity symmetry can be satisfied. Nevertheless, we show below that this minimal measurement method can be still applied to the present system with a nontrivial generalization.

In the physical Hamiltonian~\eqref{eqn:SIBlochHamiltonian1} the Pauli matrices $\sigma_{x,y,z}$ operate on the sublattice space. To complete our proof, we construct an artificial Hamiltonian which is formally equivalent to Eq.~\eqref{eqn:SIBlochHamiltonian1} $\tilde{\cal H}(\bold k)=d_x(\bold k)\tilde\sigma_x+d_y(\bold k)\tilde\sigma_y+d_z(\bold k)\tilde\sigma_z$. The only difference is that in the new Hamiltonian we assume that $\tilde\sigma_{x,y,z}$ act on a spin space which is independent of the position space. In this way, we know that the new Hamiltonian $\tilde{\cal H}(\bold k)$ is invariant under the following 2D inversion transformation on both the position and spin space
\begin{eqnarray}\label{eqn:SIinversion}
P=\tilde\sigma_z\otimes\hat R_{\rm 2D},
\end{eqnarray}
where the transformation $\hat R_{\rm 2D}$ sends $(k_{x},k_{y})\rightarrow(-k_{x},-k_{y})$. Therefore, at the four symmetric points $\{\bold\Lambda_i\}=\{(0,0),(0,\pi),(\pi,0),(\pi,\pi)\}$ the Bloch states are also parity eigenstates with $\hat P|u_{\pm}(\bold \Lambda_i)\rangle=\xi^{(\pm)}_i|u_{\pm}(\bold \Lambda_i)\rangle$, and $\xi^{(\pm)}_i=+1$ or $-1$. The topology of the artificial insulating system can be determined by the following invariant~\cite{SILiu2013}
\begin{eqnarray}\label{eqn:SIparity1}
(-1)^{\tilde\nu}=\prod_i^4\xi^{(-)}(\bold \Lambda_i)=\prod_i^4\mbox{sgn}\bigr[m_z-2(t'_A+t'_B)\cos\Lambda^i_{x'}a\cos\Lambda^i_{y'}a\bigr]=\bigr[(-1)^{\nu}\bigr]^2=1.
\end{eqnarray}
In the third line of the above equation we have defined that
\begin{eqnarray}\label{eqn:SIparity2}
(-1)^{\nu}=\prod_i^2\xi^{(-)}(\bold \Lambda_i)=\mbox{sgn}\bigr[m_z^2-4(t'_A+t'_B)^2\bigr].
\end{eqnarray}
Therefore the invariant for the constructed system $\tilde\nu\equiv0$. This indicates that the Chern number for the Hamiltonian $\tilde{\cal H}(\bold k)$ should always be even $\tilde C_1=2N$~\cite{SILiu2013}. This result is easy to understand. As pointed out previously, for the original physical system the unit cell is doubled. Accordingly, the first Brillouin zone (FBZ) of the original square lattice, denoted by $\Omega_{\rm FBZ}$ is only half of the FBZ $\tilde\Omega_{\rm FBZ}$ for the constructed artificial system. The two momenta $(k_x,k_y)$ and $(k_x+\pi,k_y+\pi)$ correspond to the same point in $\Omega_{\rm FBZ}$ (note that $\bold k$ is not the sublattice momentum, but the momentum of the original lattice system which includes both $A$ and $B$ sublattices). Therefore, the Chern number for the Bloch Hamiltonian $\tilde{\cal H}(\bold k)$ reads
\begin{eqnarray}\label{eqn:SIChern1}
\tilde C_1&=&\frac{1}{4\pi}\int_{\bold k\in\tilde\Omega_{\rm FBZ}} dk_xdk_y\bold n\cdot(\partial_{k_x}\bold n\times\partial_{k_y}\bold n)\nonumber\\
&=&\frac{1}{4\pi}\int_{\bold k\in\Omega_{\rm FBZ}} dk_xdk_y\bold n\cdot(\partial_{k_x}\bold n\times\partial_{k_y}\bold n)+\frac{1}{4\pi}\int_{\bold k\in\Omega_{\rm FBZ}+(\pi,\pi)} dk_xdk_y\bold n\cdot(\partial_{k_x}\bold n\times\partial_{k_y}\bold n)\nonumber\\
&=&\frac{1}{2\pi}\int_{\bold k\in\Omega_{\rm FBZ}} dk_xdk_y\bold n\cdot(\partial_{k_x}\bold n\times\partial_{k_y}\bold n)\nonumber\\
&=&2C_1.
\end{eqnarray}
From the number $\tilde\nu$ one cannot tell the difference of a topological phase from a trivial phase. In the next step, we shall show that the topology of the artificial system can also be determined by the invariant $\nu$ which is defined in Eq.~\eqref{eqn:SIparity2} with the parity eigenvalues at $\bold\Lambda_1=\{(0,0)$ and $\bold\Lambda_2=(0,\pi)$, half of the four parity-symmetric points in $\tilde\Omega_{\rm FBZ}$. The magnitudes $\nu=0$ and $+1$ correspond to the topologically trivial and nontrivial states, respectively. Then, together with the above relation, we can further use this invariant to characterize the topology of the original physical system.

The proof is straightforward and is valid for any two-band system satisfying the following two conditions. First, the quantum anomalous Hall phases are characterized by low Chern numbers. In particular, for the artificial system it is $\tilde C_1=\{0,\pm2\}$ and for the original physical system $C_1=\{0,\pm1\}$. Second, the system can be adiabatically connected to the one obtained under a four-fold $\hat C_4$ rotational transformation on such system. In other words, the topology is not changed under the $\hat C_4$ transformation in position and (pseudo)spin space
\begin{eqnarray}\label{eqn:SIfourfold1}
{\cal M}_4\tilde{\cal H}{\cal M}_4^{-1}\sim\tilde{\cal H}, \ {\cal M}_4=e^{i\frac{\pi}{4}\tilde\sigma_z}\otimes\hat R_{4}(\frac{\pi}{2}),
\end{eqnarray}
where $\hat R_{4}(\pi/2)$ is the $\pi/2$-rotation on the position space, transforming the Bloch momentum $(k_x,k_y)\rightarrow(k_y,-k_x)$. It is easy to see that the inversion symmetry in Eq.~\eqref{eqn:SIinversion} is given by $P={\cal M}_4^2$. By a direct check one can verify that the constructed system in our consideration belongs to the class of Hamiltonians satisfying the above conditions. What we need to prove is that the transition between a trivial phase and a topological phase must be associated with the change from $\nu=0$ to $\nu=+1$. Let the system be initially a trivial insulator. To have topological phase transition, the bulk gap should close and reopen at some momentum points. Around such momenta the bulk can be described by massive Dirac Hamiltonians, with the Dirac masses changing signs during the transition. We denote one of the Dirac momentum as $\bold k_{D1}=(k_1,k_2)$. Then, from the $\hat C_4$ symmetry we know that there are four-fold of such Dirac points $\bold k_{Dj}$ ($j=1,2,3,4$). Moreover, from the relation between the artificial and original physical systems, we have that the momentum $\bold k_{Dj}+(\pi,\pi)$ is also a Dirac point. With these results in mind, we get that when a topological phase transition occurs, the Dirac masses simultaneously reverse signs at following momenta (not necessarily independent)
\begin{eqnarray}\label{eqn:SIDirac1}
\bold k_{D1}=(k_1,k_2), \  \bold k_{D2}=(k_2,-k_1), \  \bold k_{D3}=(-k_1,-k_2), \  \bold k_{D4}=(-k_2,k_1),
\bold k_{Dj+4}=\bold k_{Dj}+(\pi,\pi), \ j=1,...,4.
\end{eqnarray}
It is easy to know that there must be even number (denoted as $2N$) of Dirac points in the above formula which are independent. On the other hand, from the symmetry we know that all these Dirac points contribute the same Chern number to the whole bulk invariant. Before and after the phase transition the Chern number changes by
\begin{eqnarray}\label{eqn:SItransition}
C_{1,\rm final}-C_{1,\rm initial}=2N.
\end{eqnarray}
We have three different cases. First, if $\bold k_{D1}$ is an inversion symmetric point, e.g. $\bold k_{D1}=\bold\Lambda_1=(0,0)$, the Eq.~\eqref{eqn:SIDirac1} includes only two independent points, $\bold k_{D1}$ and $\bold k_{D1}+(\pi,\pi)$. Second, for the case with $\bold k_{D1}=(\pi/2,\pi/2)$, the Eq.~\eqref{eqn:SIDirac1} includes four independent Dirac points, i.e. $\{\bold k_{Dj}\}=\{(\pm\pi/2,\pm\pi/2)\}$. Finally, for the rest cases all the $8$ momenta in Eq.~\eqref{eqn:SIDirac1} are independent. This implies that for the later two situations the Chern number changes by $4$ and $8$, respectively, while in the first case the Chern number changes by $2$. Therefore, for a system with low Chern number, only the first situation can happen, namely, the bulk gap must close and reopen at two inversion symmetric momenta $\bold\Lambda_1$ and $\bold\Lambda_3$ (or $\bold\Lambda_2$ and $\bold\Lambda_4$). Note that the Dirac masses at these points are equivalent to the parity eigenvalues, so the topological phase transition must be associated with the sign change of corresponding parity eigenvalues, leading to the change of the invariant $\nu$. Furthermore, it is easy to verify that the trivial phase with $\tilde C_1$ correspond to $\nu=0$, and then the topological phases with $\tilde C_1=\pm2$ are given by $\nu=+1$. Together with the relation~\eqref{eqn:SIChern1} we conclude that the invariant $\nu$ classfies the topology of the original physical system. This completes our proof.

\begin{figure}[ht]
\includegraphics[width=0.6\columnwidth]{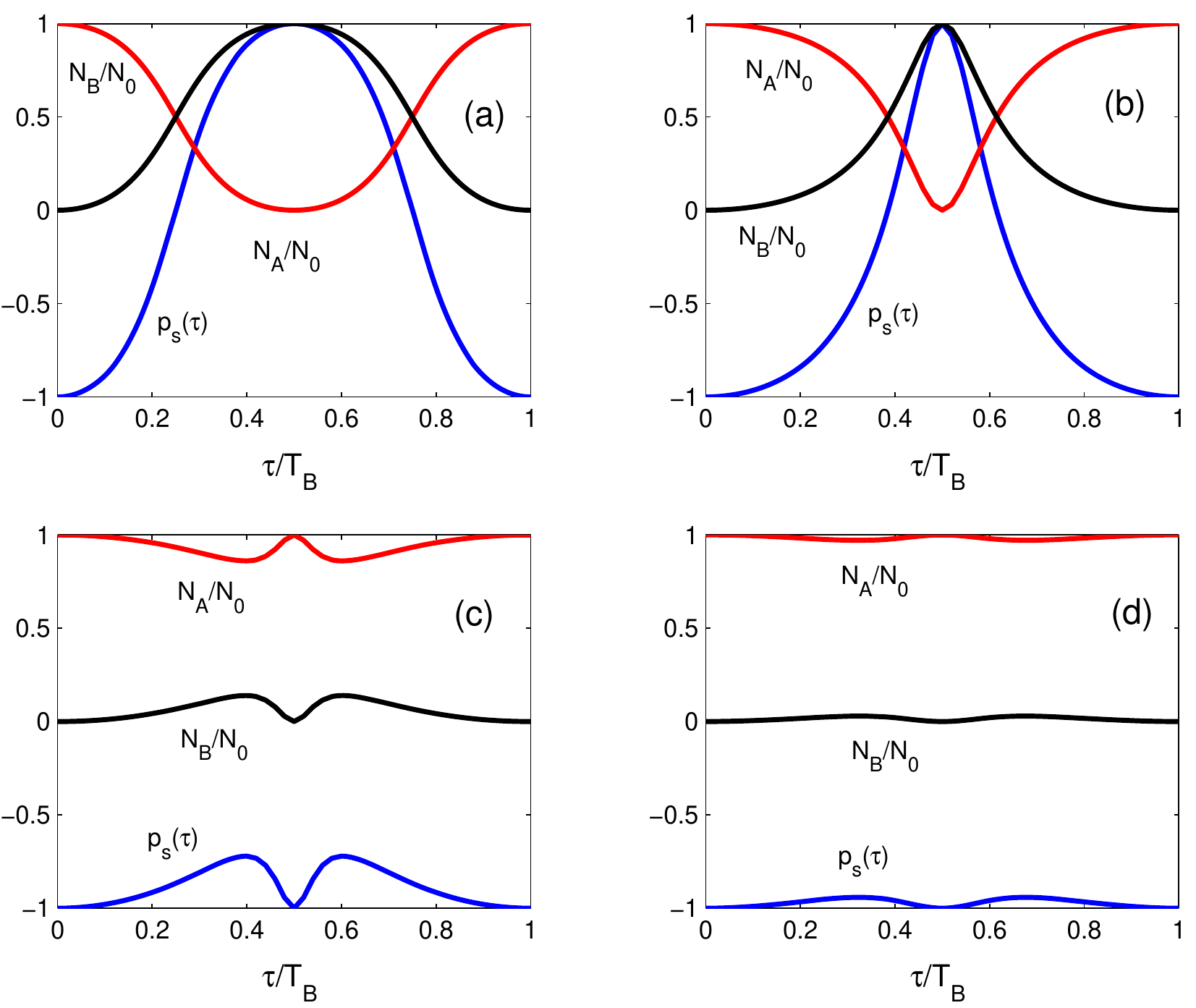}
\caption{(Color online) The time evolution of numbers of atoms populated in $A$ sites ($N_A$, in red curves) and $B$ sites ($N_B$, in black curves), and the polarization $p_s(\tau)$ (blue curves) in the Bloch oscillation, with $N_0=N_A+N_B$. (a) and (b) are topological regimes with $m_z=0$ and $m_z=-1.5(t'_A+t'_B)$, respectively; (c) and (d) are trivial regimes with $m_z=-2.5(t'_A+t'_B)$ and $m_z=-3(t'_A+t'_B)$, respectively. The hopping phase is taken as $\phi_0=\pi/4$.}\label{detection}
\end{figure}

Since the parity operator is $\sigma_z$, the parity eigenvalues are the pseudospin eigenstates. To measure the parity eigenvalues one can measure the pseudospin polarization, i.e. the population difference of atoms between $A$ and $B$ sublattices, which can be measured with {\it in situ} imaging.
The pseudospin polarization is defined by $p_s=(N_A-N_B)/(N_A+N_B)$, where $N_{A/B}$ represents the number of atoms in the $A/B$ sublattice. It follows that
\begin{eqnarray}
(-1)^{\nu}&=&\prod_{i=1}^2\mbox{sgn}\bigr[p_s(\bold\Lambda^{(i)})\bigr].
\end{eqnarray}
The topological phase corresponds to $\nu=-1$ for $|m_z|<2(t'_A+t'_B)$, and the trivial phase corresponds to $\nu=0$ for $|m_z|>2(t'_A+t'_B)$. The detection can be carried out with a pseudospin-resolved Bloch oscillation~\cite{SILiu2013}. With an external force applied along the $x$ direction, the momentum of an initial atomic cloud evolves along the direction from $\bold\Lambda_1=(0,0)$ to $\bold\Lambda_2=(0,\pi)$ (Fig.~\ref{detection}). For the topological phase, the pseudospin polarization $p_s(\tau)$ reverses sign from $\mbox{sgn}[p_s(0)]=-1$ to $\mbox{sgn}[p_s(\tau)]=+1$ at half Bloch time $\tau=T_B/2$ and returns to $\mbox{sgn}[p_s(\tau)]=-1$ at $\tau=T_B$ [Fig.\ref{detection} (a-b)]. On the other hand, in the trivial regime $|m_z|>2(t'_A+t'_B)$, the sign of the polarization keeps unchanged during the Bloch oscillation [Fig.\ref{detection} (c-d)]. Only qualitative measurements at the two symmetric momenta are needed for the experimental detection.

\subsection{Heating}

The laser beam assisted hopping between neighboring sites without spin flip can apply far-detuned Raman laser beams and therefore does no suffer from large spontaneous decay from excited states. The primary heating is induced by onsite two-photon Raman transitions which do not drive neighboring-site hopping but convert the energy difference between two Raman photons to mechanical energy of the lattice system~\cite{SIKetterle2013a,SIKetterle2013b}. Note that in the present optical Raman lattice, the Raman potentials are antisymmetric with respect to each lattice site center, the intraband scattering ($s\leftrightarrow s$ bands) is forbidden, and only the interband scattering ($s\leftrightarrow p$ bands) can heat the system. This distinguishes essentially from the conventional schemes which applies plane-wave and red-detuned Raman laser beams and have both interband and intraband onsite transitions~\cite{SIBloch2011,SIBloch2013,SIKetterle2013a,SIKetterle2013b}. To estimate the life time of the trapped cold atoms, we calculate the change rate of the mean-mechanical energy of an atom. For comparison, we consider both the present optical Raman lattice system (with the heating rate denoted as $dE_{\rm OR}/dt$) and the conventional laser-assisted schemes (denoted as $dE_{\rm CO}/dt$). Note that the $s$-$s$ band onsite transition, e.g. from an initial state with momentum $\bold k$ to the final state $\bold k'$ has the two-photon detuning $\Delta_s(\bold k,\bold k')=\delta\omega+E_{\bold k,s}-E_{\bold k',s}$, with $E_{\bold k,s}$ the $s$-band spectrum. Similarly, for $s$-$p$ band onsite transition the corresponding two-photon detuning reads $\Delta_p(\bold k,\bold k')=E_{\bold k',p}-E_{\bold k,s}-\delta\omega$, with $E_{\bold k',p}$ the energy spectrum for the $p$-band states. Let $\Omega_{ss}$ and $\Omega_{sp}$ be the two-photon Rabi-frequencies for the intraband and interband transitions, respectively. The scattering rates of the two types of transitions are given by
\begin{eqnarray}
w_{ss}(\bold k,\bold k')&=&\frac{4|\Omega_{ss}|^3}{\bigr(|\Delta_s(\bold k,\bold k')|+\sqrt{|\Delta_s(\bold k,\bold k')|^2+4|\Omega_{ss}|^2}\bigr)^2+4|\Omega_{ss}|^2}\delta(\bold k-\bold k'+\Delta\bold k),\\
w_{sp}(\bold k,\bold k')&=&\frac{4|\Omega_{sp}|^3}{\bigr(|\Delta_p(\bold k,\bold k')|+\sqrt{|\Delta_p(\bold k,\bold k')|^2+4|\Omega_{sp}|^2}\bigr)^2+4|\Omega_{sp}|^2}\delta(\bold k-\bold k'+\Delta\bold k),
\end{eqnarray}
where $\Delta\bold k$ is the momentum difference between the two Raman laser beams. The two-photon Rabi-frequencies are calculated by $\Omega_{ss}=V_R\int d^2\bold r\sin(k_1x)|\psi_{A,s}^{(0,0)}(\bold r)|^2$ and $\Omega_{sp}=V_R\int d^2\bold r\psi_{A,s}^{(0,0)}(\bold r)\cos(k_1x)\psi_{A,p}^{(0,0)}(\bold r)$, with $\psi^{(\vec j)}_{A,p}$ the local $p$-orbital wave function. We then obtain the rates of change of the mean-mechanical energy
by
\begin{eqnarray}\label{eqn:SIrate}
\frac{dE_{\rm OR}}{dt}&=&\frac{1}{N}\sum_{\bold k,\bold k'}w_{sp}(\bold k,\bold k')\delta\omega=\Gamma_{\rm OR},\\
\frac{dE_{\rm CO}}{dt}&=&\frac{1}{N}\biggr[\sum_{\bold k,\bold k'}w_{ss}(\bold k,\bold k')+\sum_{\bold k,\bold k'}w_{sp}(\bold k,\bold k')\biggr]\delta\omega=\Gamma_{\rm CO},
\end{eqnarray}
where $N$ is the number of lattice sites.
In the above formulae we have denoted by $\Gamma_{\rm OR}$ and $\Gamma_{\rm CO}$ the heating rates of the optical Raman lattice system and the conventional lattice systems, respectively. Typically the bandwidths of $s$ and $p$ bands are much less than the $s$-$p$ band gap which is $E_{sp}=2(V_0E_R)^{1/2}$. We then approximate that $\Delta_s(\bold k,\bold k')\approx\delta\omega$ and $\Delta_p(\bold k,\bold k')\approx E_{sp}-\delta\omega$, and
\begin{eqnarray}\label{eqn:SIrate}
\frac{dE_{\rm OR}}{dt}&\approx&\frac{4|\Omega_{sp}|^3}{\bigr(2|E_{sp}-\delta\omega|+2|\Omega_{sp}|^2/|E_{sp}-\delta\omega|\bigr)^2+4|\Omega_{sp}|^2}\delta\omega,\\
\frac{dE_{\rm CO}}{dt}&\approx&\frac{4|\Omega_{ss}|^3}{\bigr(2|\delta\omega|+2|\Omega_{ss}|^2/|\delta\omega|\bigr)^2+4|\Omega_{ss}|^2}\delta\omega+\nonumber\\
&&+\frac{4|\Omega_{sp}|^3}{\bigr(2|E_{sp}-\delta\omega|+2|\Omega_{sp}|^2/|E_{sp}-\delta\omega|\bigr)^2+4|\Omega_{sp}|^2}\delta\omega.
\end{eqnarray}
For the parameter with $V_0=5E_R$, one can verify that $\Omega_{sp}\approx0.5\Omega_{ss}$. Substituting this result into the equation of $dE_{\rm CO}/dt$ we solve numerically that the minimum heating rate for $\Gamma_{\rm CO}$ is obtained by setting $\delta\omega\approx0.4E_{sp}$. Namely, for the conventional laser-assisted-hopping schemes, the minimum heating (denoted as $\Gamma^{\rm min}_{\rm CO}$) requires that the frequency difference between Raman lasers be close to the half of the $s$-$p$ band gap~\cite{SIKetterle2013a,SIKetterle2013b}. With these results in mind we obtain directly from the above two equations the following relation
\begin{eqnarray}\label{eqn:SIlifetime}
\Gamma_{\rm OR}\simeq\frac{1}{16}\frac{\Delta}{E_{sp}-\Delta}\Gamma^{\rm min}_{\rm CO}.
\end{eqnarray}
Here we have taken $\delta\omega\approx\Delta$ for the formula of $\Gamma_{\rm OR}$. It is easy to see that for the present optical Raman lattice, a relatively small on-site energy offset is preferred. One can set the typical parameters that $\Delta\ll E_{sp}$, and we have then $\Gamma_{\rm OR}\ll\Gamma^{\rm min}_{\rm CO}$, which shows that the present optical Raman lattice has much less heating than that in the conventional schemes. In particular, with the parameter regime that $V_0/E_R=4, \tilde V_0/E_R=1$, and $\delta\omega=\Delta=0.6E_R$, one finds that $\Gamma_{\rm OR}\simeq0.03\Gamma^{\rm min}_{\rm CO}$. Furthermore, for $^{87}$Rb atoms using $\lambda=532$nm and for $V_R=0.5E_R$ the life time of the optical Raman lattice $\tau=V_0/\Gamma_{\rm OR}\simeq1.67$s, which is extremely long enough for realistic experiments.

\section{Chiral spin liquid phase}

\subsection{$J_1$-$J_2$-$K$ model}

In this section we proceed to study the realization of the $J_1$-$J_2$-$K$ model. For this we consider a spin-$1/2$ two-copy version of the QAH model together with repulsive Fermi Hubbard interaction, as illustrated in Fig.~\ref{fig:phasedgm} (a). In the single-particle regime each spin species forms a QAH insulator with the same Chern number. In the presence of Hubbard interaction, we have the total Hamiltonian that
$H=\sum_{\bold k}\hat {\mathcal C}^{\dag}(\bold
k)\mathcal{H}_0(\bold k)\hat {\mathcal C}(\bold k)+H_{\rm int}$ with $\hat
{\mathcal C}(\bold k)=(\hat c_{a\uparrow}(\bold k), \hat c_{b\uparrow}(\bold k),\hat c_{a\downarrow}(\bold k), \hat c_{b\downarrow}(\bold k))^T$ and obtain
\begin{eqnarray}\label{eqn:SIinteraction1}
\mathcal{H}_0(\bold k)&=&\sum_{\alpha=x,y,z}d_{\alpha}(\bold k)\sigma_\alpha\otimes I,\\
H_{\rm int}&=&\sum_iUn_{i\uparrow}n_{i\downarrow},
\end{eqnarray}
where $U$ is the strength of Hubbard interaction. For simplicity we take that $m_z=0$. In the large-$U$ regime, the double occupancy of each site will be fully suppressed, and the system always becomes a Mott insulator.
We can derive an effective spin model by only considering the Hilbert space with single occupancy and treating the hopping term $\mathcal H_0$ as perturbations.
\[
\langle \{\sigma\} |H_{\rm eff}|\{\sigma'\}\rangle = \sum_n { \langle{ \{\sigma} \}| \mathcal H_0^n |\{\sigma'\}\rangle \over U^{n-1}} ,
\]
where $ |\{\sigma\}\rangle$ and $ |\{\sigma'\}\rangle$ are two different spin configurations with only single occupancy.
Note that to reflect the broken time-reversal symmetry in the spin-model, we should at least keep the terms up to third order of $t_0/U,t'_\mu/U$, which gives the three-spin interactions through triangular loops.
Counting in all spin configurations and up to the third order perturbation we can reach the following effective Hamiltonian for the spin degree of freedom
\begin{eqnarray}\label{eqn:spinmodel1}
H_{\rm eff}=\sum_{\langle i,j\rangle}J_1\bold S_i\cdot\bold S_j+\sum_{\langle\langle i,j\rangle\rangle}J_2\bold S_i\cdot\bold S_j+\sum_{i,j,k\in\bigtriangleup}K\sin(\phi_{ijk})\bold S_i\cdot(\bold S_j\times\bold S_k),
\end{eqnarray}
where $J_1=4t_0^2/U, J_2=4t_1^2/U$ with $t_1=t'_A\approx t'_B$, $K=24t_0^2t'_1/U^2$, and $\phi_{ijk}$ is the Aharonov-Bohm phase acquired through a closed triangular loop $i\rightarrow j\rightarrow k\rightarrow i$. The spin operators are defined by $S_{i,z}=(c^\dag_{i,\uparrow}c_{i,\uparrow}-c^\dag_{i,\downarrow}c_{i,\downarrow})/2$, $S_{i,x}=(c^\dag_{i,\uparrow}c_{i,\downarrow}+c^\dag_{i,\downarrow}c_{i,\uparrow})/2$, and $S_{i,y}=i(c^\dag_{i,\uparrow}c_{i,\downarrow}-c^\dag_{i,\downarrow}c_{i,\uparrow})/2$. It is clear that the third order $K$-term emerges only when the time-reversal symmetry is broken in the system. The summation in the third term means that each set of $(i,j,k)$ consists of a minimum triangular. It can be verified that $\phi_{ijk}=\pi/2$ when $\phi_0=\pi/4$. In this case all spins experience a uniform $U(1)$ magnetic field and the spin system respects the translational symmetry which, however, is not respected by the original free fermion system. We note that the next-order coupling are four-spin interacting terms, with the four spins located in the four sites of a plaquette. Since the flux across each plaquette is $\pi$, the four-spin interacting terms do not break time-reversal symmetry, and are expected to have much weaker effect on the chiral spin liquid phase. The magnitudes of $J_{1,2}$ are fully controllable by tuning $t_{1,2}$ through $\tilde V_0$ and Raman potentials.

\subsection{Chiral spin liquid phase}

We solve the spinon mean-field phase diagram for the present $J_1$-$J_2$-$K$ model with $\phi_{ijk}=\pi/2$. This model contains at least three phases.
First, when $J_1$ dominates, the system is unfrustrated and it supports a Neel anti-ferromagnetic order [Fig.~\ref{fig:phasedgm} (b)]. Secondly, when $J_2$ dominates, the system is also unfrustrated and can have a stripe anti-ferromagnetic order, in which case the staggered spin order exists only in the $x$ or $y$ direction [Fig.~\ref{fig:phasedgm} (c)].  Finally, when $K$ is large enough, the system prefers a chiral spin liquid state~\cite{SIKalmeyer,SIWen1989}.


The different phases can be studied using trial (mean field) wave function method. We introduce the anyonic spinons $f_i=(f_{i\uparrow}, f_{i\downarrow})^T$ to represent the spin operators as $\pmb S_i = f_i^\dag{\pmb \sigma\over2} f_i$ under the particle number constraint $f_i^\dag f_i=1$. In the mean field theory for $U(1)$ spin liquid, the spin interactions can be rewritten as the following,
\begin{eqnarray}
&&\pmb S_i\cdot \pmb S_j = -{1\over2}\hat\chi_{ij} \hat\chi_{ji},\\
&&\pmb S_i\times \pmb S_j\cdot \pmb S_k = {1\over2i}{1\over6} \{ \left[\hat\chi_{ij}\hat\chi_{jk}\hat\chi_{ki} + {\rm cyclic}(ijk)\right] - h.c.\},
\end{eqnarray}
where $\hat\chi_{ij} = \hat\chi_{ji}^\dag = f_i^\dag f_j$
is the spinon hopping operator. Here we do not consider the spinon pairings since our interest is mainly focused on the $U(1)$ chiral spin liquid phases. In general the spinon hopping term is complex, and the spin chirality term can give rise to a phase $e^{i\phi_\Delta}$, with $\phi_\Delta={\rm Arg}(\langle\hat\chi_{ji}\rangle\langle\hat\chi_{kj}\rangle\langle\hat\chi_{ik}\rangle)$ the flux experience by spinons after hopping through a close triangular loop. For chiral spin liquid state, in this work we have numerically verified that the ground state corresponds to $\phi_\Delta={\pi\over2}$. Actually, with a Landau gauge choice for the mean field theory one can confirm that the $\pi/2$-flux state in each triangle has the lowest energy.
Therefore, for convenience we introduce the trial mean field parameters
\begin{eqnarray*}
\chi_1&=&\langle \hat\chi_{ii+1_x}\rangle^*|_{x_i={\rm odd}}=-\langle \hat\chi_{ii+1_x}\rangle|_{x_i={\rm even}}\\
&=& \langle \hat\chi_{ii+1_y}\rangle|_{y_i={\rm odd}}=-\langle \hat\chi_{ii+1_y}\rangle|^*_{y_i={\rm even}},\\
\chi_2&=&\langle \hat\chi_{ii+1_x+1_y}\rangle=\chi_2^*
\end{eqnarray*}
to decouple the spin interactions. Here $\chi_2$ is assumed to be real and the hopping phase is carried by $\chi_1$ for the chiral spin liquid phase. With other gauge choice for the mean field parameters we shall get the same phase diagram. It is easiy to see that with the above trial mean field parameters the spinons experience a uniform $U(1)$ gauge field, with the magnetic flux through each triangular being $\pi/2$ and through each plaquette being $\pi$~\cite{SIWen1989}.

Furthermore, since the system may contains symmetry breaking orders at some parameter region, we also need to introduce the following magnetization order parameters to describe the Neel order and stripe order, respectively
\begin{eqnarray}
M_n=(-1)^{x_i+y_i}\langle S^z_i\rangle, \ M_s = (-1)^{x_i}\langle S^z_i\rangle.
\end{eqnarray}
Now we can decouple the spin interactions by these (trial) mean field parameters as
\begin{eqnarray*}
\pmb S_i\cdot \pmb S_{i+1_x} |_{x_i={\rm odd}}&=& (-{1\over2}\chi_1 \hat\chi_{ii+1_x} + h.c) + {1\over2}|\chi_1|^2+(-1)^{x_i+y_i} (M_n S^z_{i+1_x} - M_n S^z_i) + M_n^2,\\
\pmb S_i\cdot \pmb S_{i+1_x} |_{x_i={\rm even}}&=& -(-{1\over2}\chi_1^* \hat\chi_{ii+1_x} + h.c) + {1\over2}|\chi_1|^2 +(-1)^{x_i+y_i} (M_n S^z_{i+1_x} - M_n S^z_i) + M_n^2,\\
\pmb S_i\cdot \pmb S_{i+1_y} |_{y_i={\rm odd}}&=& (-{1\over2}\chi_1^* \hat\chi_{ii+1_y} + h.c) + {1\over2}|\chi_1|^2+(-1)^{x_i+y_i} (M_n S^z_{i+1_x} - M_n S^z_i) + M_n^2,\\
\pmb S_i\cdot \pmb S_{i+1_y} |_{y_i={\rm even}}&=& -(-{1\over2}\chi_1 \hat\chi_{ii+1_y} + h.c) + {1\over2}|\chi_1|^2 +(-1)^{x_i+y_i} (M_n S^z_{i+1_x} - M_n S^z_i) + M_n^2,\\
\pmb S_i\cdot \pmb S_{i+1_x+1_y} &=& (-{1\over2}\chi_2 \hat\chi_{ii+1_x+1_y} + h.c) + {1\over2}|\chi_2|^2 + (-1)^{x_i+y_i} (M_n S^z_{i+1_x+y}+M_n S^z_i) - M_n^2\\
&&+ (-1)^{x_i}(M_s S^z_{i+1_x+1_y}-M_s S^z_i) + M_s^2,\\
\pmb S_i\times \pmb S_j\cdot \pmb S_k&=& {1\over4i}[\langle \hat\chi_{ij}\hat\chi_{jk}\rangle\hat\chi_{ki} + {\rm cyclic}(ijk) - h.c.]-2{1\over4i}[|\chi_1^2\chi_2|e^{-i\phi_\Delta}-h.c.]\nonumber\\
&=& {1\over4i}[{|\chi_1^2\chi_2|e^{-i\phi_\Delta}\over\langle\hat\chi_{ki}\rangle}\hat\chi_{ki} + {\rm cyclic}(ijk) - h.c.]
-2{1\over4i}[|\chi_1^2\chi_2|e^{-i\phi_\Delta}-h.c.].\nonumber
\end{eqnarray*}
Notice that both the Neel order $M_n$ and the stripe order $M_s$ are collinear, so they don't appear in decoupling
the spin chirality interaction term $\pmb S_i\times \pmb S_j\cdot \pmb S_k$. In the chiral spin liquid phase $M_n=M_s=0$, and from the above expressions we find that the spinons experience a uniform magnetic field which leads to the quantum Hall effect for spin degree of freedom. Then the ground state of the chiral spin liquid phase is captured by the bosonic $\nu=1/2$ Laughlin state~\cite{SILaughlin} which has chiral gapless anyonic spinon excitations in the edge~\cite{SIKalmeyer,SIWen1989}.

\begin{figure}[ht]
\includegraphics[width=0.8\columnwidth]{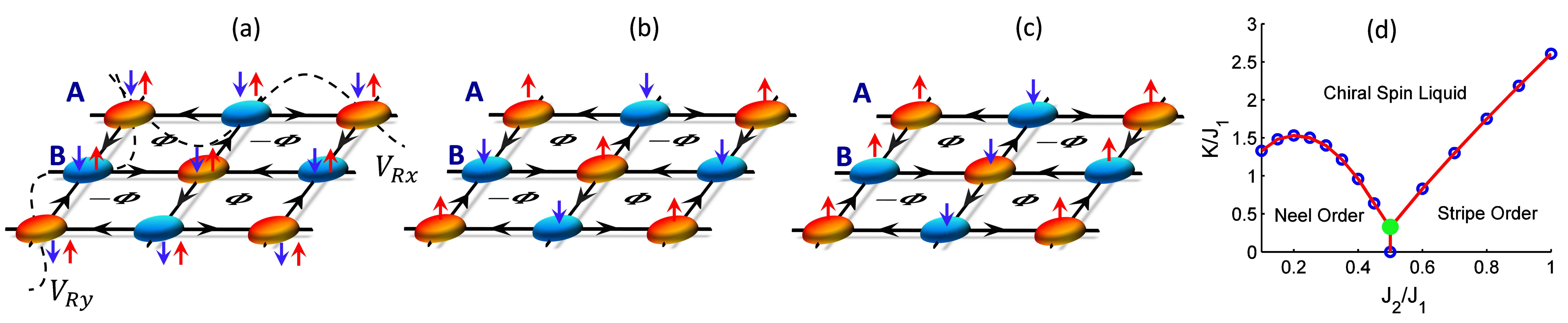}
\caption{(Color online) (a) Sketch of the spin-$1/2$ two-copy version of the quantum anomalous Hall system; (b) The antiferromagnetic (Neel) order; (c) The stripe order; (d) The phase diagram for the $J_1$-$J_2$-$K$ model.} \label{fig:SIphasedgm}
\end{figure}
The numerical results are shown in Fig.~\ref{fig:SIphasedgm} (d). It can be read that the three different phases are clearly dominated by the different coupling terms in the Hamiltonian~\eqref{eqn:spinmodel1}. The antiferromagnetic (Neel) or stripe order is obtained when the nearest- ($J_1$) or next-nearest-neighbor ($J_2$) coupling term dominates. On the other hand, when the three-spin interactions ($K$-terms) dominate, the chiral spin liquid phase results. In this case, no local order exists and the spin degree of freedom exhibits a gap in the bulk, while supports chiral gapless spinons in the edge~\cite{SIKalmeyer,SIWen1989}. It is seen that for $K=0$, the transition between Neel and stripe orders occurs at $J_2=J_1/2$, at which point the system becomes most frustrated, and then increasing $K$ can soon drive and stabilize the chiral spin liquid phase. In particular, the chiral spin liquid phase appears above the point $J_1=2J_2\approx3.2K$ [green point in Fig.~\ref{fig:SIphasedgm} (d)]. If taking that $V_R=0.52E_R$ and $U=8t_0$, we find that $J_1\approx2J_2\approx2\pi\times19$Hz and $K\approx0.53J_1$, which belongs to the chiral spin liquid phase region.


\noindent

\end{document}